\documentclass[%
10pt,
prb,
aps,
twocolumn,
amsmath,
amssymb,
reprint,%
showpacs,
showkeys,
]{revtex4-1}
\pdfoutput=1

\usepackage[USenglish]{babel}
\usepackage[T1]{fontenc}
\usepackage[ansinew]{inputenc}
\usepackage{lmodern} %Type1-font for non-english texts and characters
\usepackage{amsmath}
\usepackage{amsfonts}
\usepackage{amssymb}
\usepackage{graphicx}
\usepackage{amsthm}

\usepackage[protrusion=true,expansion=true]{microtype}				% Better typography

\usepackage{booktabs}																					
\usepackage{array}
\usepackage{multirow}

\usepackage{hyperref}

\newcommand{\Eexp}[1]{ \ensuremath{ \textrm{e}^{ #1} } }

\newcommand{\BoldVec}[1]{ \ensuremath{ {\boldsymbol{#1}} } }

\newcommand{\NN}[2]{ \ensuremath{ {\left\langle #1,#2 \right\rangle} } }
\newcommand{\NumberOp}[2]{ \ensuremath{ { #1 _{#2}^{\dagger} #1_{#2} } }}

\newcommand{\Abs}[1]{ \ensuremath{ { \left| #1 \right| } } }
\newcommand{\Hopping}[4]{ \ensuremath{ {#1} _{#2}^{\dagger} {#3} _{#4} } }

\newcommand{\BCSTermUp}[4]{\ensuremath{ { #1 _{#2 \uparrow }^{\dagger} #3 _{#4 \downarrow }^{\dagger}} }}
\newcommand{\BCSTermDown}[4]{\ensuremath{ { #1 _{#2 \downarrow }^{\dagger} #3 _{#4 \uparrow }^{\dagger}} }}

\begin{document}
\title{Defects in the $\left(d+id\right)$-wave superconducting state in heavily doped graphene}
\author{Tomas Löthman}
\author{Annica M. Black-Schaffer}
\affiliation{Department of Physics and Astronomy, Uppsala University, Box 516, S-751 20 Uppsala, Sweden}
\date{\today}
%\today
\pacs{%
	74.70.Wz, % Superconducting materials, carbon-based materials,
	81.05.ue, % Carbon-based materials, Graphene
	73.22.Pr, % condensed matter, graphene		
	74.20.Mn, %Superconductivity, nonconventional mechanisms,
	74.20.Rp, %Pairing symmetries (superconductivity),
	74.62.Dh % Impurities effects on superconductivity; Doping effects on transition temperature (superconductivity); Defects, crystal, superconductors
}

\keywords{Graphene, Superconductivity, Impurities}

\begin{abstract}
	A chiral time-reversal symmetry breaking $(d+id)$-wave superconducting state is likely to emerge in graphene doped close to the Van Hove singularity. As heavy doping procedures are expected to introduce defects, we here investigate the effects of microscopic defects on the $(d+id)$-wave superconducting state at the Van Hove singularity. We find that, while the superconducting order is reduced near a defect, the $(d+id)$-wave state remains intact and recovers in an exponential manner away from the defect. The recovery length is found to be on the order of one lattice constant for weak couplings. This is comparable to the recovery length of a conventional $s$-wave state, demonstrating that the unconventional $(d+id)$-wave state is quite resilient to defects. Moreover, we find no significant changes between a single site defect and more extended defects, such as a bivacancy. While the $(d+id)$-wave state is fully gapped, we also show that defects introduce localized midgap states with non-zero energies, which should be accessible via scanning probe experiments.	
\end{abstract}

\maketitle

%================================================
%================================================
%============Introduction========================
%================================================
%================================================
\section{Introduction}
	Graphene consists of a single layer of carbon atoms arranged in a honeycomb lattice, and it was first isolated from flakes of graphite in 2004.\cite{Novoselov22102004} Undoped graphene has a vanishing density of states (DOS) at the Fermi energy and a linear quasiparticle energy dispersion relationship $\epsilon_{\BoldVec{k}} = v_F | \BoldVec{k} |$. \cite{RevModPhys.81.109} The vanishing DOS makes undoped graphene rather unsusceptible to interaction-driven instabilities, although the role of interactions in graphene are still being investigated. \cite{PhysRevB.63.134421, RevModPhys.84.1067}
	
	As graphene is electron or hole doped the DOS at the Fermi energy increases and the increased DOS has been shown to render graphene susceptible to several interaction-driven instabilities; such as spin-density-wave \cite{PhysRevB.84.125404} (SDW), charge-density-wave \cite{PhysRevLett.100.146404} (CDW), Pomeranchuk \cite{1367-2630-10-11-113009}, or superconducting instabilities \cite{PhysRevB.75.134512,PhysRevB.78.205431,PhysRevB.81.085431,ScBosonExchange, PhysRevLett.100.146404}. In particular, the DOS is logarithmically divergent at the Van Hove singularity (VHS), which is located at a $\pi$-band filling of $3/8$ and $5/8$.\cite{RevModPhys.81.109} Electron densities approaching the VHS doping have experimentally been reported using chemical doping \cite{PhysRevLett.104.136803} and electrolytic gating. \cite{PhysRevLett.105.256805} 
	
	Both perturbative \cite{ChiralSuperconductivity} and functional \cite{PhysRevB.85.035414, PhysRevB.86.020507} renormalization group calculations, which take into account competing orders, have shown that near, or at, the VHS doping an unconventional chiral $(d+id)$-wave superconducting state is likely to emerge from repulsive interactions. This is consistent with the findings of strong coupling approaches,\cite{PhysRevB.75.134512,PhysRevB.84.121410,PhysRevB.81.085431} as well as results based on the Kohn-Luttinger mechanism.\cite{PhysRevB.78.205431} Superconductivity may also persist to the lightly doped case, although with increased competition from other instabilities. \cite{PhysRevB.87.094521,PhysRevLett.100.146404}
	
	The chiral $(d+id)$-wave state is a topological superconducting state with several unusual properties, such as spontaneous edge currents and Majorana fermions at finite magnetic fields.\cite{PhysRevLett.109.197001} The symmetry of the honeycomb lattice automatically makes the $d$-wave superconducting channel twofold degenerate, and this in turn leads to the stabilization of the time-reversal symmetry breaking $(d+id)$-wave combination below the superconducting transition temperature.\cite{PhysRevB.78.205431,PhysRevB.75.134512} Moreover, it has been proposed that the intrinsic $(d+id)$-wave state is enhanced by superconducting proximity effect in graphene Josephson junctions with $d$-wave superconducting contacts,\cite{PhysRevB.81.014517} as well as in the core of doubly quantized vortices in conventional $s$-wave superconducting graphene.\cite{PhysRevB.88.104506}
	
	However, any method intended to introduce the doping required to reach the $(d+id)$-wave state in graphene is also likely to introduce a significant amount of defects to the graphene sheet, which might weaken, or possibly even destroy, the superconducting state. It is thus highly relevant to understand the effects of defects on the unconventional, time-reversal symmetry breaking $(d+id)$-wave state. In this article we, therefore, investigate the effects of microscopic defects on the $(d+id)$-wave state in graphene doped to the VHS. In doing so, we ask how disruptive defects are to the $(d+id)$-wave state, how quickly the state heals, and whether there are any qualitative changes brought about by the defects. We also compare the results for the unconventional $(d+id)$-wave state to that of a conventional $s$-wave state, since conventional superconductors are known to be robust against any non-magnetic disorder as they are protected by the Anderson theorem. \cite{RevModPhys.78.373,Anderson195926}
		
	We here study both vacancies and charge neutral impurities, which are both representative microscopic defects. We find that the superconducting order is reduced near the defects, but that the $(d+id)$-wave state remains intact and recovers in an exponential manner away from the defects. The recovery length is found to be on the order of one lattice constant for weak couplings, which is comparable to the recovery length found for a conventional $s$-wave state. This suggests that the $(d+id)$-wave state is in fact quite resilient to defects despite its unconventional and exotic nature. We also find no notable difference between single vacancies and bivacancies, demonstrating that even breaking the point symmetry group does not significantly perturb the $(d+id)$-wave state. In addition, while the $(d+id)$-wave state is completely gapped, we show that defects introduce localized midgap states with non-zero excitation energies, which should be accessible via scanning probe experiments.	

\section{Method} \label{sec:Method}
	To model the electronic properties of graphene we use a simple nearest-neighbor hopping Hamiltonian for the $\pi$-bands which emerge from the out-of-plane $p_z$-orbitals; that is, 
	\begin{equation} \label{eqn:HoppingH}
		\hat{H}_{0} 
		=
		-
		t 
		\sum_{\NN{i}{j} , \sigma}{ 
			\big( 
				\Hopping{a}{i \sigma}{b}{j \sigma} 
				+ \textrm{H.c}
			\big)
		}
		+ \mu 
		\sum_{i , \sigma}{
			\big( 
				\NumberOp{a}{i \sigma} 
				+
				\NumberOp{b}{i \sigma}
			\big)
		}
		,
	\end{equation}
	where $i$ and $j$ are site indices, $a$ and $b$ refers to respective sublattice, $\NN{i}{j}$ indicates that the sum is over nearest-neighbors, $\sigma$ denotes the spin, $t$ is the hopping amplitude, and $\mu$ is the chemical potential; for undoped graphene it is known that $t \approx 2.5-2.9 \: \textrm{eV}$, and the VHS is located at $\mu = \pm t $. \cite{RevModPhys.81.109}
	
	It has been shown that the real space pairing of the $(d+id)$-wave state near the VHS is very localized, which is consistent with the large DOS at the VHS strongly screening long-range interactions. \cite{PhysRevB.86.020507,PhysRevB.85.035414} We therefore mainly consider superconducting paring localized to nearest-neighbor bonds. Nearest-neighbor and next-nearest-neighbor pairing were both considered in Ref.~[\onlinecite{PhysRevLett.109.197001}], where it was found that the properties of the superconducting state are insensitive to the exact pairing location. To investigate the properties of superconductivity in graphene we thus add a general mean-field nearest-neighbor bond spin-singlet pairing term:
	\begin{align} \label{eqn:HNNPairing}
		\hat{H}_{\Delta}
		=
		-
		\sum_{\NN{i}{j}}{
			\left[
				\Delta_{ij}
				\big(
					\BCSTermUp{a}{i}{b}{j}
					-
					\BCSTermDown{a}{i}{b}{j}
				\big)
				+
				\textrm{H.c}
			\right]
		}
		+
		\frac{1}{J} 
		\sum_{\NN{i}{j}}{
			\Abs{\Delta_{ij}}^2
		},
	\end{align}
	 where $J$ denotes the superconducting coupling strength. Minimizing the free-energy with respect to the parameters $\Delta_{ij}$ yields the self-consistency equations,
	\begin{equation} \label{eq:NNDeltaSCEq}
		\Delta_{ij}
		=
		\frac{J}{2}
		\langle
			\BCSTermUp{a}{i}{b}{j}
			-
			\BCSTermDown{a}{i}{b}{j}
		\rangle
		.
	\end{equation}
	The full Hamiltonian  $\hat{H} = \hat{H}_{0}  + \hat{H}_{\Delta}$ was also obtained in Ref.~[\onlinecite{PhysRevB.75.134512}] via a mean-field decoupling of a phenomenological resonant-valence-bond (RVB) model, which had been introduced in an earlier context to describe graphene-like systems. \cite{PhysRevB.65.212505, TheFiveFoldWay} 
	
	For comparison we also consider a conventional $s$-wave superconducting state, which can simplest be modeled by mean-field decoupling of an attractive on-site Hubbard model on the honeycomb lattice: 
		\begin{align} \label{eqn:HHub}
		\hat{H}_{U}
		=
		&
		-
		\sum_{i}{
			\big( 
				\Delta^{a}_{i}
				\BCSTermUp{a}{i}{a}{i}
				+
				\Delta^{b}_{i}
				\BCSTermUp{b}{i}{b}{i}
				+
				\textrm{H.c}
			\big)
		}
		\nonumber
		\\
		&
		+ \frac{1}{U} 
		\sum_{i}{
			\big(
				\Abs{\Delta^{a}_{i}}^2
				+
				\Abs{\Delta^{b}_{i}}^2
			\big)
		}
		,
	\end{align}
	where $U$ corresponds to the on-site potential strength. Minimizing the free-energy with respect to the order parameters, $\Delta^{a,b}_{i}$, yields the self-consistency equations,
	\begin{equation} \label{eq:NNDeltaSCEq}
		\Delta^{a}_{i}
		=
		\frac{U}{2}
		\langle
			\BCSTermUp{a}{i}{a}{i}
		\rangle
		,
	\end{equation}
	with an analogous expression for the $b$-sites. 
	
	The mean-field treatment used in this work neglects fluctuations and  quantitative results may change when going beyond mean-field; for instance, the transition temperature may be modified. \cite{ScBosonExchange} However, we do not expect the qualitative effects of defects found in this work to be modified.

	\subsection{Defect-free Graphene}
	In the absence of defects we will assume that the order parameters, $\Delta_{ij}$, are translationally invariant, while allowing for directional dependence as is indicative of an unconventional superconducting state; that is, $\Delta_{ij} \mapsto \Delta_{\BoldVec{\xi}} $, where $\BoldVec{\xi}$ labels the three inequivalent bond directions. With these assumptions the Hamiltonian is translationally invariant. It is therefore block-diagonal in reciprocal space, and it can be diagonalized by a Bogoliubov-Valatin transformation. We find it convenient to gather the remaining free parameters into a vector, $\vec{\Delta} = \big( \Delta_{\BoldVec{\xi}_1}, \Delta_{\BoldVec{\xi}_2}, \Delta_{\BoldVec{\xi}_3} \big)^{T}$, which we now regard as the order parameter. 	 
	
	Near the transition temperature, $T_c$, the self-consistency equations can be linearized. The linearized self-consistency equations are invariant with respect to the hexagonal, $D_{6h}$, point symmetry group of the honeycomb lattice, and the solutions are, therefore, classified according to symmetry and belong to specific irreducible representations of $D_{6h}$. It has been found that for $\hat{H}$ there is one extended $s$-wave solution, which belongs to the identity representation $A_{1g}$, and a twofold degenerate $d$-wave solution space, which belongs to $E_{2g}$.\cite{PhysRevB.75.134512} The basis set for the solutions can thus be taken to be
	\begin{gather} \label{eqn:BasisSet}
	\vec{\Delta}_{s} =
		\frac{1}{\sqrt{3}}
		\begin{pmatrix} 
			1	\\
			1	\\
			1
		\end{pmatrix},
		\quad
		%%%%%%%%%%%%%%
		\vec{\Delta}_{d_{x^2-y^2}} =
		\frac{1}{\sqrt{6}} 
		\begin{pmatrix} 
			2				\\
			-1		\\
			-1
		\end{pmatrix},
		%%%%%%%%%%%%%%%
		\\ \nonumber
		\vec{\Delta}_{d_{xy}} =
		\frac{1}{\sqrt{2}}
		\begin{pmatrix} 
			0		\\
			-1	\\
			1
		\end{pmatrix}
		,
	\end{gather}
	where $\vec{\Delta}_{s}$ form a basis of $A_{1g}$, and $\vec{\Delta}_{d_{x^2-y^2}}$ and $\vec{\Delta}_{d_{xy}}$ form a basis of $E_{2g}$.
	We here refer to the two solutions $\vec{\Delta}_{d_{x^2-y^2}}$ and $\vec{\Delta}_{d_{xy}}$, as $d_{x^2-y^2}$-wave, and $d_{xy}$-wave, respectively, since for these two solutions the pairing has the symmetry of the representative functions $x^2-y^2$ and $xy$. Below $T_c$ the self-consistency equation becomes non-linear and admixtures of the different solutions are in general allowed. It has been shown that one of the $\left(d\pm id\right)$-wave states are, as we shall also see below, preferred below $T_c$ for not too strong coupling strengths, $J$, and a doping up to and around the VHS, \cite{PhysRevB.75.134512} where
	\[
		\vec{\Delta}_{d_{x^2-y^2} \mp i d_{xy}} =
		\frac{1}{\sqrt{3}}
		\begin{pmatrix} 
			1																\\
			\Eexp{\pm \frac{2 i \pi }{3}}		\\
			\Eexp{\mp \frac{2 i \pi }{3}}
		\end{pmatrix}
		.
	\]
		
	The magnitude of the order parameter, $ | \vec{\Delta} | $, can be used as a measure of the strength of the superconductivity. For the $(d+id)$-wave state it is approximately true that $T_c \propto | \vec{\Delta} |$, where $T_c$ is the mean-field transition temperature. Numerically we find that $T_c \approx 0.52 | \vec{\Delta} |$. This in turn implies that $ 2 | \vec{\Delta} | / T_c \approx 3.85$, which is within $10 \%$ of the $2 \pi / \Eexp{\gamma} \approx 3.5 $ value for an isotropic BCS model.
	
	\subsection{Introduction of Defects}
	To investigate the effects of microscopic defects on the superconductivity we either remove one site, as to form a vacancy, or we add an on-site potential to a site, to model a charge neutral impurity. We consider also a bivacancy for which two adjacent sites are removed. The defect is then repeated periodically after a given number of unit cells as to form a supercell in order to use periodic boundary conditions. In the presence of defects the order parameter becomes position dependent. To make connection with the defect-free case we also introduce a local wave-character. We define the $x$-wave character as $ | \vec{\Delta}  \cdot \vec{\Delta_{x}} | / |\vec{\Delta}|$, where $\vec{\Delta_{x}}$ is the $x$-wave solution basis, e.g.\ $\vec{\Delta}_{s}$, $\vec{\Delta}_{d_{x^2-y^2}}$, $\vec{\Delta}_{d_{xy}}$, or $\vec{\Delta}_{d_{x^2-y^2} \pm i d_{xy}}$.
	
	\subsection{Numerical Methods}
	The self-consistent values of the order parameter, $\vec{\Delta}$, were found by numerical minimization of the free-energy, and all calculations were started from random complex values of the order parameters. Both the termination tolerance on the order parameters and on the free-energy were set to values less than $10^{-12}$, as to ensure convergence in the phases. The $\BoldVec{k}$-point convergence was checked for the defect-free case with $J/t=0.5$, $\mu/t=1$, and $T=0$, and for a $80 \times 80$ $\BoldVec{k}$-point sampling the calculations have by and large converged; nonetheless, larger samplings were mostly used. For the supercell calculations a $\BoldVec{k}$-point sampling yielding as many, or more than, the corresponding number of eigenvalues were used. The validity of this rule was checked for a few test cases by both increasing and reducing the $\BoldVec{k}$-point sampling.
	
	%%%%%%%%%%%%%%%%%%%%%%%%%%%%%%%%%%%%%%%
	%%%%%%%%%%%%%%%%%%%%%%%%%%%%%%%%%%%%%%%
	%%%%%%%%%%%%%%  Results  %%%%%%%%%%%%%%
	%%%%%%%%%%%%%%%%%%%%%%%%%%%%%%%%%%%%%%%
	%%%%%%%%%%%%%%%%%%%%%%%%%%%%%%%%%%%%%%%
	\section{Results}
	
	\subsection{Low Temperature Phase Diagram}
	
	\begin{figure*}
		\includegraphics[width=0.94\textwidth]{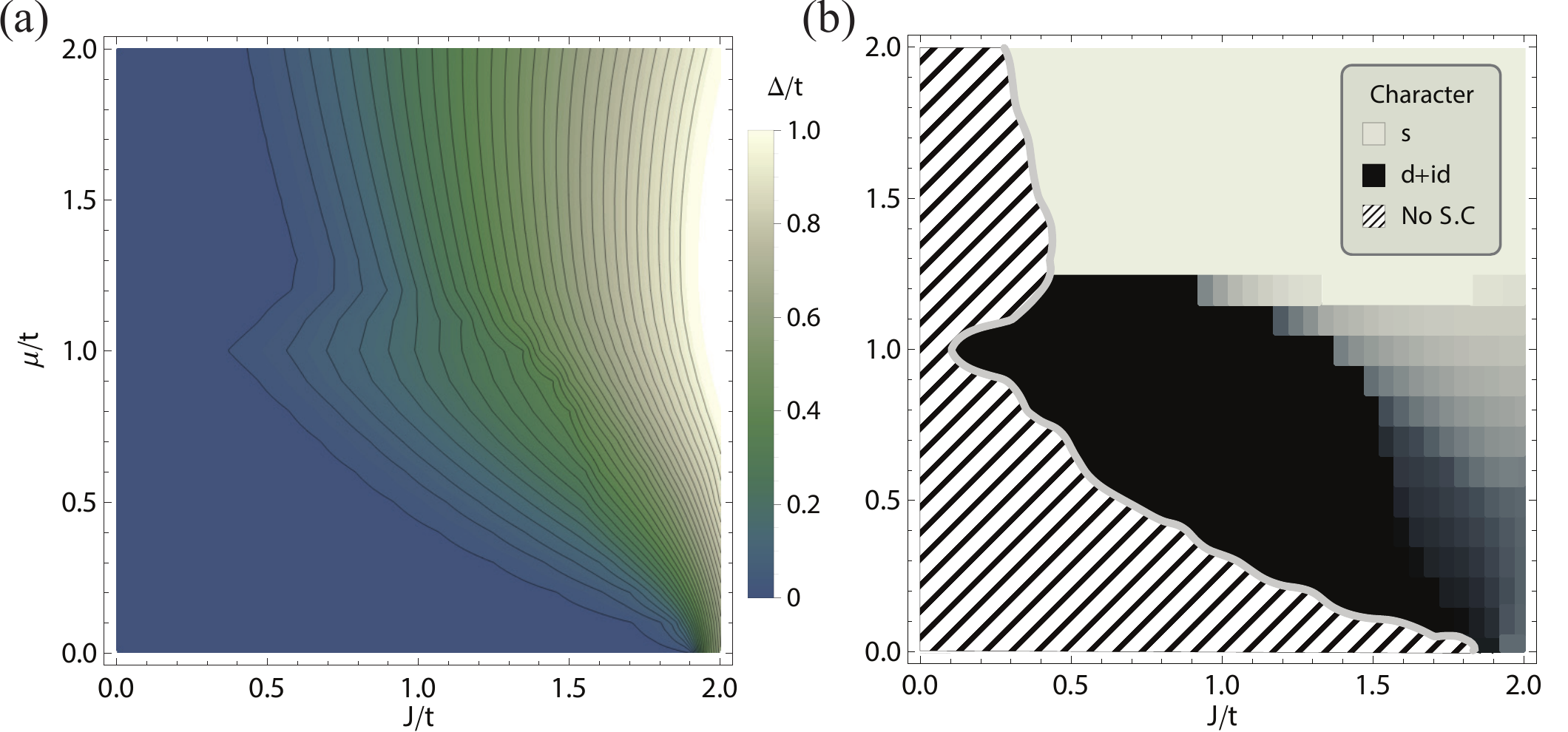}
		\caption{
			(Color online) (a) Contour plot of the order parameter magnitude, $| \vec{\Delta} | / t $, in units of the hopping amplitude $t$ at $T=0$, and as a function of the dimensionless chemical potential, $\mu/t$, and the interaction strength, $J/t$. The distance between the contours is $0.032$, there are $30$ contours, and the first contour is at $0.016$. %
			(b) Wave-character of the order parameter. The $s$-wave character is shown in a bright shade and the $(d + id)$-wave is shown in a dark shade. The white hatched region indicates that there is no superconductivity or that the order parameter is too small to be accurately classified. The grey regions indicate that the order parameter is not pure $s$-wave or pure $(d + id)$-wave, but, rather, a superposition of these.
		}
		\label{fig:Phase}
	\end{figure*}
	
	We first consider the mean-field phase diagram of a defect-free graphene sheet over a large doping and coupling constant regime. Figure \ref{fig:Phase}(a) shows the order parameter magnitude, $| \vec{\Delta} | / t$, as a function of the two dimensionless parameters: the chemical potential, $\mu/t$, and the interaction strength, $J/t$. We see that undoped graphene is not superconducting unless the interaction strength, $J/t$, surpasses a large critical value.\cite{PhysRevB.75.134512} Superconductivity is, however, enhanced at the VHS doping, $\mu/t=1$, where it remains appreciable even for weak interactions which is consistent with the renormalization group results.\cite{PhysRevB.86.020507,ChiralSuperconductivity,PhysRevB.87.094521} Fig.\ \ref{fig:Phase}(b) shows the wave-character of the order parameter as a function of the chemical potential, $\mu/t$, and the interaction strength, $J/t$, at zero temperature. It is seen that the ridge extending from the VHS is pure $\vec{\Delta}_{d_{x^2-y^2} + i d_{xy}}$ (or $\vec{\Delta}_{d_{x^2-y^2} - i d_{xy}}$), but that there is a transition into a general admixture with a $s$-wave component for interaction strengths which are larger than $J/t \approx 1.4$, whereas for doping levels far beyond the VHS the state is pure $s$-wave.
	
	\subsection{Lattice Vacancy}
\begin{figure}
		\centering
			\includegraphics[width=0.42\textwidth]{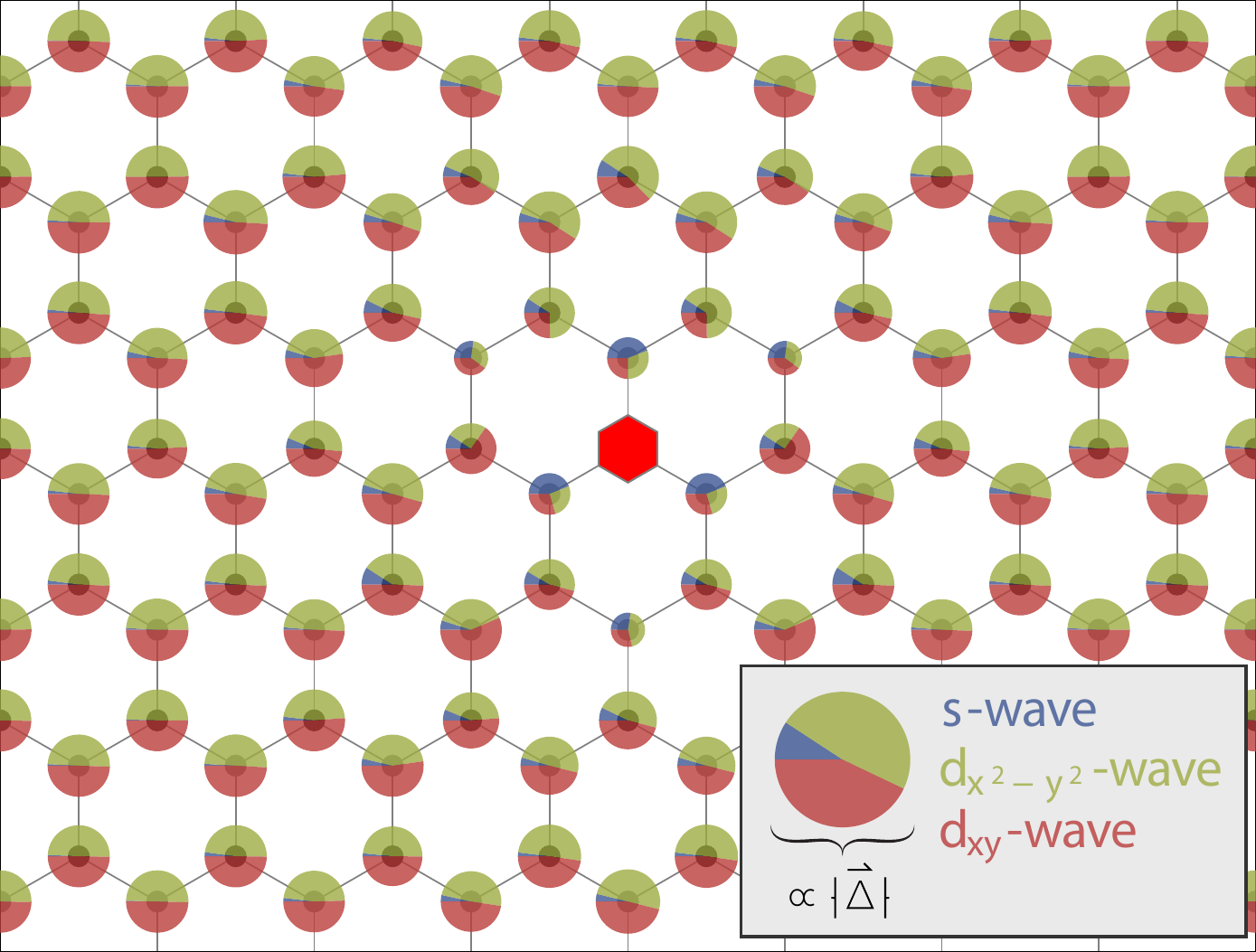}
		\caption{
			(Color online) A qualitative view of the order parameter, $\vec{\Delta}$, near a vacancy (red polygon) for $J/t = 0.875$, $\mu/t = 1$ at $T=0$. The local wave-character of each site is shown by a pie chart; the segments of which are proportional to respective wave-character. The radius of each pie chart is proportional to the magnitude of the order parameter at the site.
		}	
		\label{fig:Pie}
	\end{figure}

	We now turn to the effects brought about by the introduction of vacancies. To investigate this we remove the first $a$-site from each supercell. The investigation of vacancies captures also the essential features of strong charge neutral impurities, as both disrupt only a local site in the lattice. Figure \ref{fig:Pie} gives a qualitative depiction of changes to the order parameter, $\vec{\Delta}$, in the vicinity of a vacancy. At each site a pie chart shows the local wave-character, and the radius of the pie chart is proportional to  $|\vec{\Delta}|$. It is seen that at the vacancy the order parameter adapts to the conditions imposed by the vacancy, and that the magnitude is reduced in the vicinity of the vacancy. The $(d + id)$-wave state is, nonetheless, seen to recover at a distance of $2$ -- $3$ lattice constants away from the vacancy. Moreover, we find that the state remains close to the $(d + id)$-wave state down to a $5 \times 5$ supercell for a wide range of interaction strengths. We thus conclude that superconductivity is disrupted only locally, and that it converges back to the $(d + id)$-wave state without qualitatively altering the state even for high concentrations of vacancies. 
	
\begin{figure}
		\centering
			\includegraphics[width=0.45\textwidth]{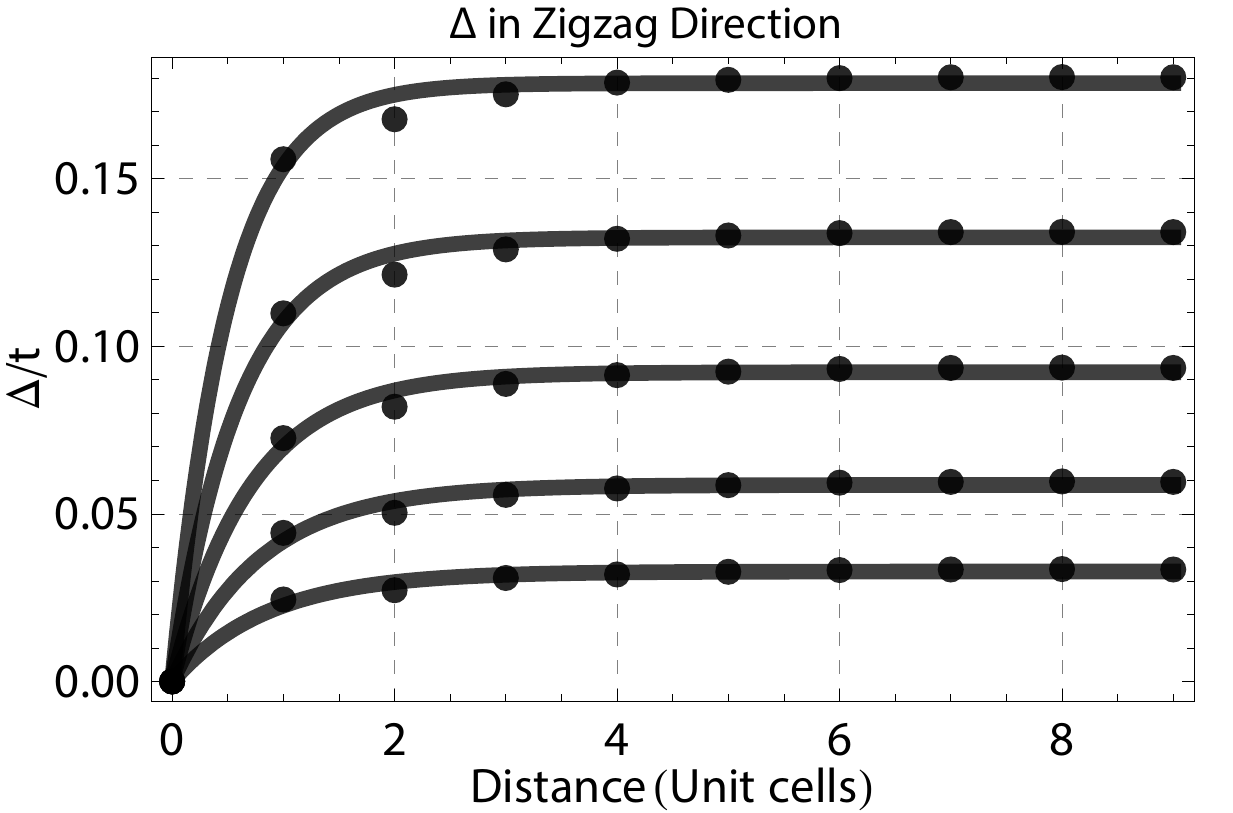}
		\caption{
			The order parameter magnitude, $ | \vec{\Delta} |$, (black dots) as a function of the distance from the lattice vacancy, and for several values of the interaction strength, $J/t = 0.5$, $0.625$, $0.75$, $0.875$, $1.0$ (increasing with $ | \vec{\Delta} |$). To each data set the function $A [1 - \exp \! \left( - x / \xi \right) ]$ is least-square-fitted (solid lines), where $x$ is the distance from the vacancy, and $\xi$ is the recovery length. 
		}	
		\label{fig:Fit}
	\end{figure}

	\paragraph{Recovery length.} 
	To more precisely measure the effects of vacancies, we single out for investigation the order parameters in one of the most affected directions (zigzag). We consider, for this direction, the norms, $ | \vec{\Delta} |$, for multiple values of the interaction strength, $J/t$, and for a large $16 \times 16$ supercell that ensures that the vacancies are sufficiently isolated. This allows us to see how the order parameter recovers, and how fast the recovery is. The numerical values are shown in Fig.\ \ref{fig:Fit}, and for each interaction strength an exponential recovery function of the form, $A [1 - \exp \! \left( - x / \xi \right) ]$, has been least-square-fitted to the data points. The convergence back to the defect-free graphene superconducting state is seen to be quite well described by this exponential function form.
	
\begin{figure}
		\centering
			\includegraphics[width=0.45\textwidth]{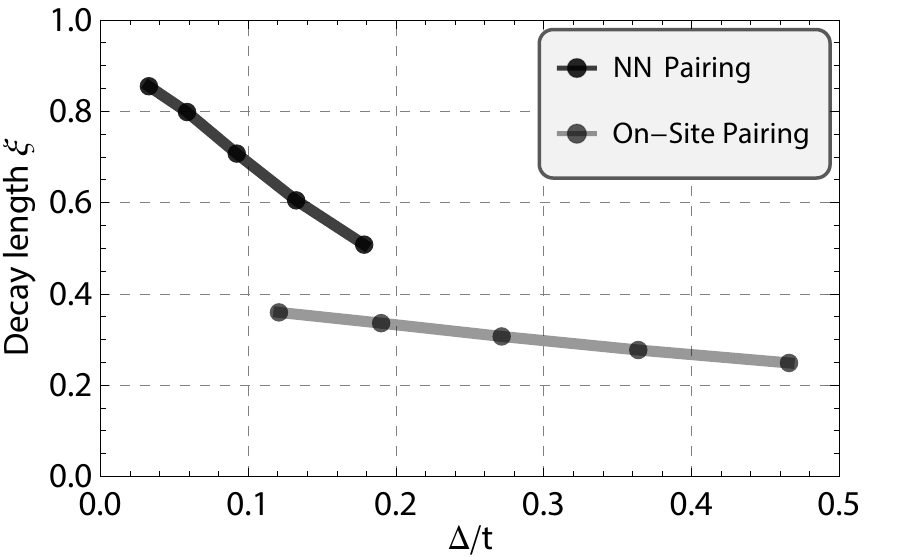}
		\caption{
			Comparison of the recovery length, $\xi$, for the $(d+id)$-wave and a conventional $s$-wave state, where $\xi$ is given in units of the lattice constant $a$ and as a function of the defect-free value of the order parameter magnitude, $| \vec{\Delta} | / t$. The $(d+id)$-wave data points are the same as those in Fig.~\ref{fig:Fit}, whereas the $s$-wave data points corresponds to $U/t = 1.25$, $1.5$, $0.875$, $2.0$, and $2.25$ (increasing with $| \vec{\Delta} | / t$).
		}
		\label{fig:AttHubCompVac}
	\end{figure}

	The recovery length, $\xi$, is plotted in Fig.\ \ref{fig:AttHubCompVac} as a function of the defect-free value of the order parameter magnitude, $| \vec{\Delta} | / t$. For comparison the corresponding recovery lengths for a conventional $s$-wave superconducting state on the honeycomb lattice are also shown in Fig.\ \ref{fig:AttHubCompVac}. It is clear from Fig.\ \ref{fig:AttHubCompVac} that the dependence of the recovery length on the order parameter magnitude is approximately linear for both pairing symmetries. The recovery length of the $(d + id)$-wave state is seen to be somewhat more sensitive to the order parameter magnitude than is the recovery length of the $s$-wave pairing, although this result is also somewhat dependent on the implementation of the $(d+id)$ state on e.g.~nearest or next-nearest neighbor bonds. When extrapolated to weak coupling strengths the unconventional $(d+id)$-wave state is seen to have $\xi/a \sim 1$ for both nearest and next-nearest neighbor (not shown) pairing, where $a$ is the lattice constant, whereas conventional $s$-wave pairing has $\xi/a \sim 0.4$. The recovery lengths for both the $s$- and $(d+id)$-wave superconducting states are, therefore, comparable, even when extrapolated to the weak pairing regime, despite the unconventional nature of the $(d+id)$-wave state. 
	
	A semi-quantitative interpretation of the recovery length, $\xi$, is that the order parameter fully heals within a length of $\sim 3 \xi$. We therefore expect that two vacancies will start to interact appreciably when they are within twice this distance; that is, when they are within a distance of $\sim 6 \xi$ of each other. This is consistent with our findings that the state remains close to the $(d + id)$-wave state down to a $5 \times 5$ supercell for a wide range of interaction strengths.
	
\begin{figure}
		\centering
			\includegraphics[width=0.45\textwidth]{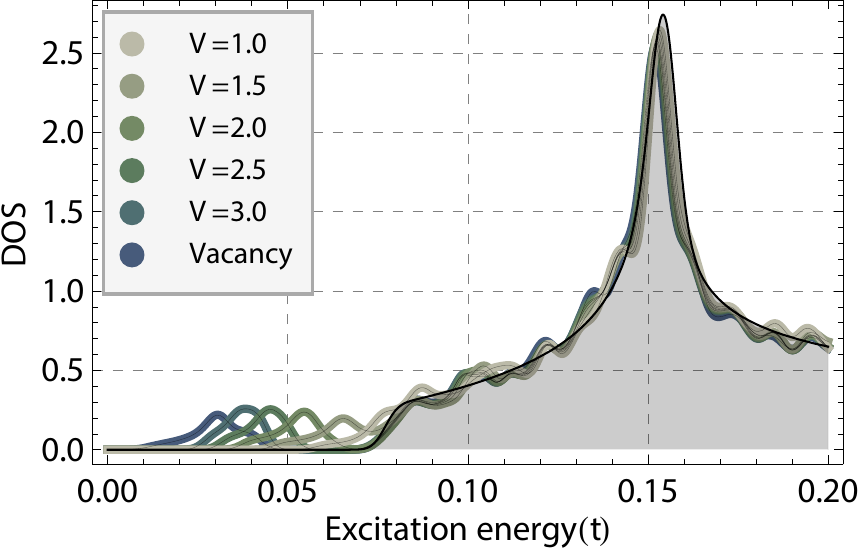}
		\caption{
			(Color online) The DOS in units of states per unit cell and $t$ of a $12 \times 12$ supercell for increasing on-site impurity potential strengths, $V$, for $J/t=0.875$, $\mu/t=1.0$, and $T=0$. The gray shaded area bounded by the solid black line indicates the defect-free DOS. The densities of states are plotted with a Gaussian smearing of width $25t \cdot 10^{-4}$.
		}
		\label{fig:DOSImp}
	\end{figure}

	\paragraph{Density of states.}
	We next turn to the changes in the density of states (DOS) as a result of the introduction of  defects. The DOS determines most thermodynamic quantities, and it is accessible via photoemission spectroscopy or scanning probe experiments. 
	For a defect free graphene sheet the $(d + id)$-wave state is fully gapped at the Fermi surface for all finite doping levels. However, as seen in Fig.\ \ref{fig:DOSImp} a vacancy or impurity introduce midgap states. The defect-free DOS has, for the parameters used in Fig.\ \ref{fig:DOSImp}, a full gap energy gap up to about $0.07t$ and a coherence peak at about $0.15t$. By studying the progression of increasing impurity strengths, $V$, the midgap states are seen to emerge from the band edge and gradually drift towards the gap center although never reaching zero energy.
		
\begin{figure}
		\centering
			\includegraphics[width=0.45\textwidth]{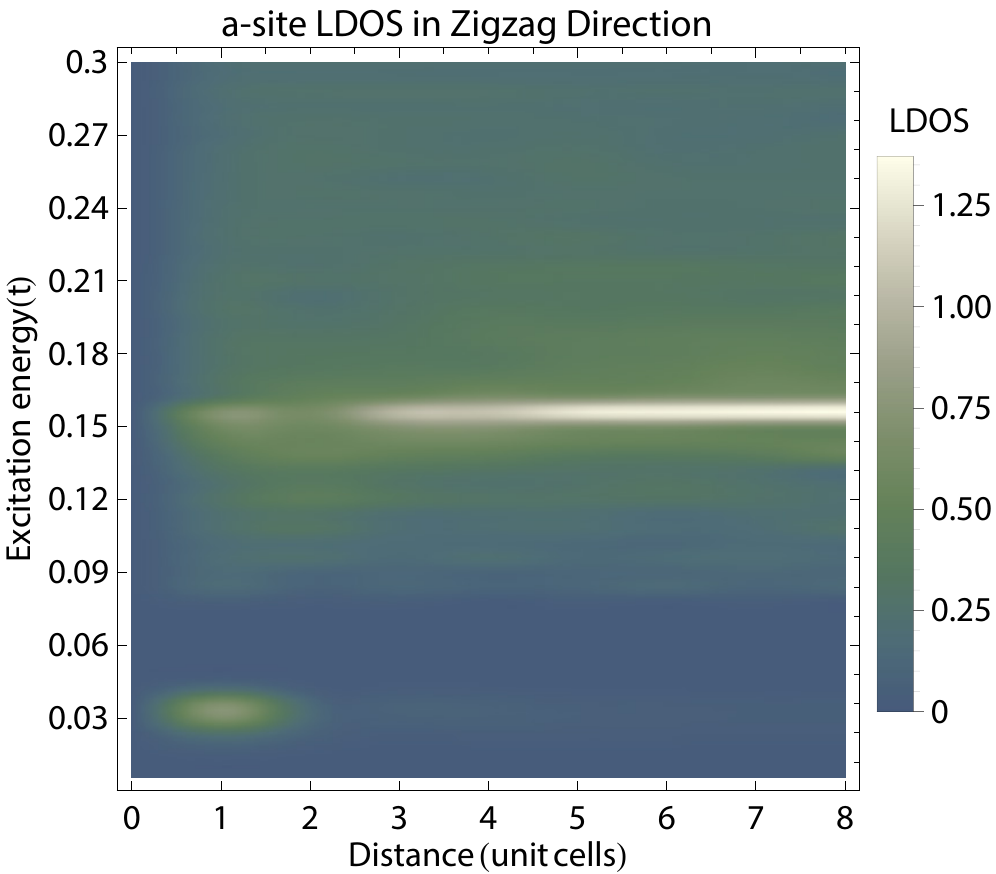}
		\caption{
			(Color online) Low energy LDOS as a function of the distance from a vacancy for a $16 \times 16$ supercell, $J/t=0.875$ and $\mu/t=1.0$ at $T=0$. Plotted with a $2$nd order interpolation and a bin counting width of $ 6t \cdot 10^{-3} $.
		}
		\label{fig:LDOS}
	\end{figure}

	With modern scanning tunneling microscopy and spectroscopy it should also be possible to access the local changes introduced by point defects. \cite{RevModPhys.78.373} In particular, the local density of states (LDOS) should be accessible via differential current measurements. Fig.\ \ref{fig:LDOS} shows the energy resolved, low energy, $a$-site LDOS as a function of distance from the vacancy in the most affected direction (zigzag). The midgap states are seen to be very localized to the vacancy; nonetheless, the energy gap itself remains largely unaltered.
	
\begin{figure}
		\centering
			\includegraphics[width=0.45\textwidth]{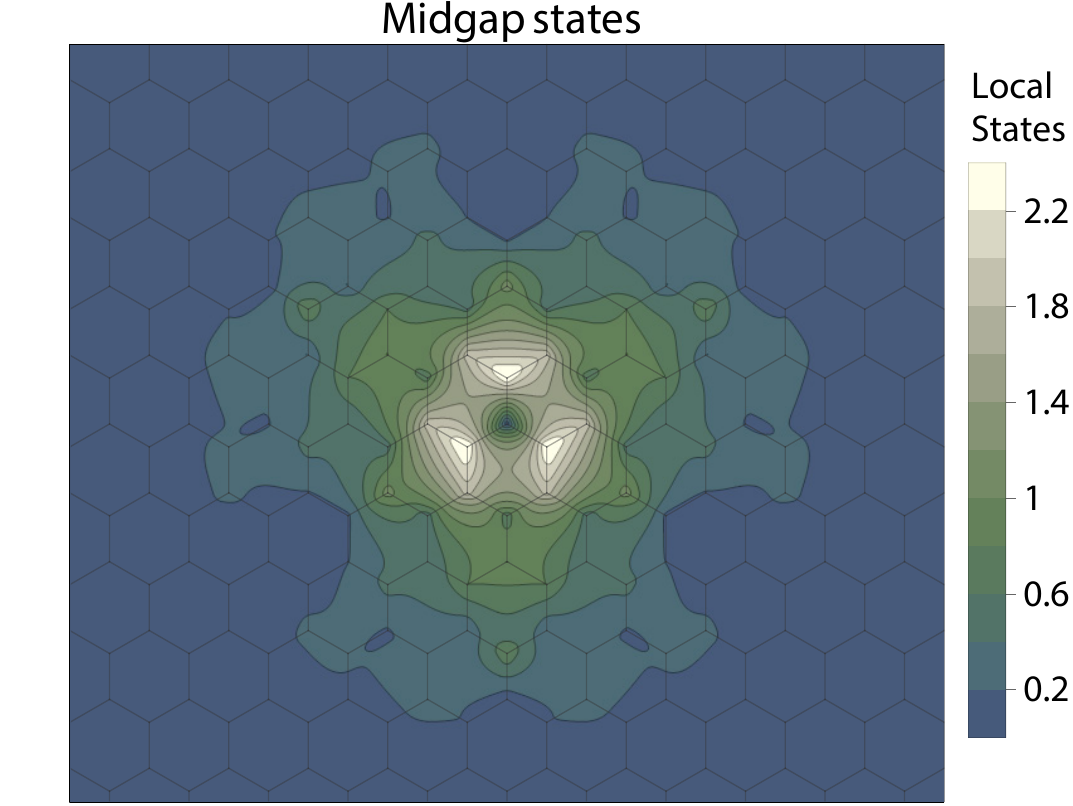}
		\caption{
			(Color online) Contour plot of the LDOS near a vacancy (located at the center of the figure) integrated up to the energy gap on each site for a $16 \times 16$ supercell, $J/t=0.875$ and $\mu/t=1.0$ at $T=0$. Plotted using a linear interpolation.
		}
		\label{fig:MidGap}
	\end{figure}

	The midgap states should also be accessible via constant bias tunneling measurements. Fig.\ \ref{fig:MidGap} shows the in-gap, integrated LDOS for each site; that is, the quantity shown is the LDOS integrated up to the defect-free gap energy. From Fig.\ \ref{fig:MidGap} it is clear that the midgap states display the symmetry of the lattice and are very localized around the vacancies. This also explains the very localized effects of single impurities on the superconducting state seen in Fig.~\ref{fig:Pie}.	
	
	\subsection{Bivacancy}
\begin{figure}
		\centering
			\includegraphics[width=0.42\textwidth]{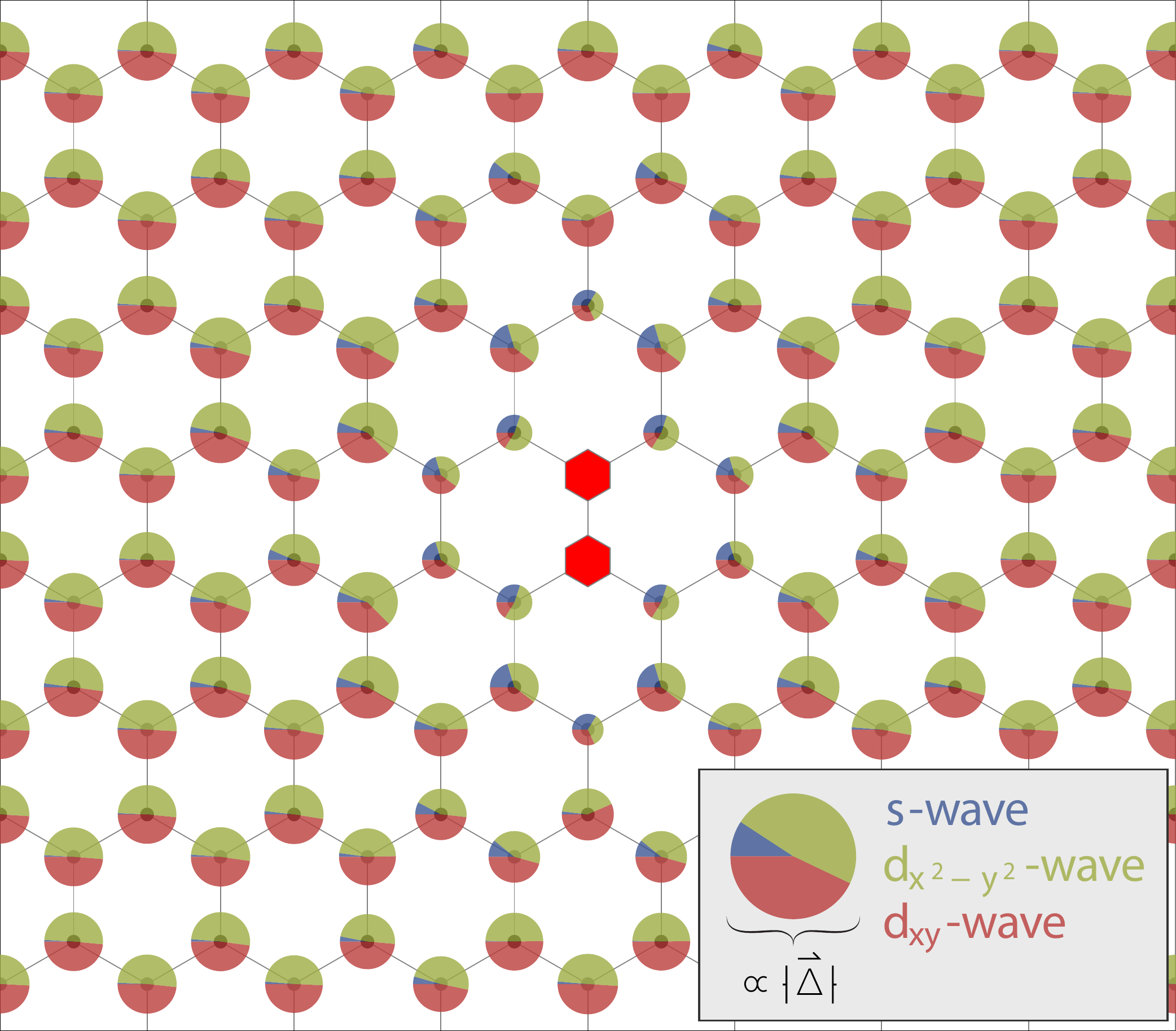}
		\caption{
			(Color online) A qualitative view of the order parameter, $\vec{\Delta}$, near a bivacancy (red polygons) for $J/t = 0.875$ and $\mu/t = 1$ at $T=0$. The local wave-character of each site is shown by a pie chart; the segments of which are proportional to respective wave-character. The radius of each pie chart is proportional to the magnitude of the order parameter at the site.
		}	
		\label{fig:PieBiVac}
	\end{figure}

	At a single-site vacancy or impurity the point symmetry is not reduced. One might therefore ask whether the effects of an extended defect are considerably different from that of a single site defect. This is of particular relevance for an unconventional superconducting state such as the $(d+id)$-wave state which exhibits a directional dependence. In order to investigate this we consider a bivacancy that breaks the sixfold point symmetry of the honeycomb lattice. Fig.\ \ref{fig:PieBiVac} gives a qualitative view of the order parameter in the same manner as was done for a single vacancy in Fig.\ \ref{fig:Pie}. While the two cases are different, there are no considerable qualitative changes in the disruption to the order parameter. In the vicinity of the bivacancy the order parameter magnitude is reduced, and the wave-character adapts to the constraints imposed by the defect. The order is, however, seen to heal back to the bulk $(d+id)$-wave state at a distance of $2$ -- $3$ lattice constants away from the defect, as was also found for a single vacancy; that is, it remains true that $\xi / a \sim 1$. Moreover, the healing length, $\xi$, has the same dependence on the defect-free order parameter magnitude, $| \vec{\Delta} |$, as for the single vacancy shown in Fig.\ \ref{fig:AttHubCompVac}; even if it is, as to be expected, slightly larger in the bivacancy case. Our findings for a single vacancy thus extend to the case of a bivacancy, and we conclude that the $(d+id)$-wave state is not any more sensitive to impurities breaking the sixfold rotational lattice symmetry than it is to single-site impurities.
		
%%%%%%%%%%%%%%%%%%%%%%%%%%%%%%%%%%%%%%%%%
%%%%%%%%%%%%%%%%%%%%%%%%%%%%%%%%%%%%%%%%%
%%%%%%%%  Summary and Conclusions  %%%%%%
%%%%%%%%%%%%%%%%%%%%%%%%%%%%%%%%%%%%%%%%%
%%%%%%%%%%%%%%%%%%%%%%%%%%%%%%%%%%%%%%%%%

\section{Summary and Conclusions}
In summary we have investigated the effects of defects on the chiral time-reversal symmetry breaking $(d+id)$-wave state proposed to appear in graphene doped close to the VHS.\cite{PhysRevB.86.020507,ChiralSuperconductivity,PhysRevB.87.094521} We have found that, despite its unconventional nature, the $(d+id)$-wave state is quite robust against vacancies and impurities, and it remains intact for both point and more extended, rotational symmetry breaking defects, such as a bivacancy. Away from a defect the superconducting order parameter recovers quickly to the bulk $(d+id)$-wave state with a healing length that is only about one lattice constant for weak couplings. This is comparable to that of a conventional $s$-wave superconducting state on the honeycomb lattice, demonstrating that the $(d+id)$-wave state is
quite resilient to defects. Our results are thus very promising for an experimental discovery of the $(d+id)$-wave state as inducing heavy doping into graphene will undoubtedly result in a certain level of imperfections in the graphene sheet. We have also found that vacancies and impurities introduce a set of midgap states into the fully gapped $(d+id)$-wave state, which should be accessible via scanning probe experiments. These states emerge from the band edge and gradually drift towards the gap center with increasing impurity strength, although they never reach zero energy even for vacancies. We furthermore find that the midgap states are very localized to the impurity, which is consistent with the limited effect of the impurity found on the overall superconducting state.

\begin{acknowledgments}
	We thank Kristofer Björnson for helpful discussions and the Swedish Research Council (VR) for support.
\end{acknowledgments}


\begin{thebibliography}{26}%
\makeatletter
\providecommand \@ifxundefined [1]{%
 \@ifx{#1\undefined}
}%
\providecommand \@ifnum [1]{%
 \ifnum #1\expandafter \@firstoftwo
 \else \expandafter \@secondoftwo
 \fi
}%
\providecommand \@ifx [1]{%
 \ifx #1\expandafter \@firstoftwo
 \else \expandafter \@secondoftwo
 \fi
}%
\providecommand \natexlab [1]{#1}%
\providecommand \enquote  [1]{``#1''}%
\providecommand \bibnamefont  [1]{#1}%
\providecommand \bibfnamefont [1]{#1}%
\providecommand \citenamefont [1]{#1}%
\providecommand \href@noop [0]{\@secondoftwo}%
\providecommand \href [0]{\begingroup \@sanitize@url \@href}%
\providecommand \@href[1]{\@@startlink{#1}\@@href}%
\providecommand \@@href[1]{\endgroup#1\@@endlink}%
\providecommand \@sanitize@url [0]{\catcode `\\12\catcode `\$12\catcode
  `\&12\catcode `\#12\catcode `\^12\catcode `\_12\catcode `\%12\relax}%
\providecommand \@@startlink[1]{}%
\providecommand \@@endlink[0]{}%
\providecommand \url  [0]{\begingroup\@sanitize@url \@url }%
\providecommand \@url [1]{\endgroup\@href {#1}{\urlprefix }}%
\providecommand \urlprefix  [0]{URL }%
\providecommand \Eprint [0]{\href }%
\@ifxundefined \urlstyle {%
  \providecommand \doi  [0]{\begingroup \@sanitize@url \@doi}%
  \providecommand \@doi [1]{\endgroup \@@startlink {\doibase
  #1}doi:\discretionary {}{}{}#1\@@endlink }%
}{%
  \providecommand \doi  [0]{doi:\discretionary{}{}{}\begingroup
  \urlstyle{rm}\Url }%
}%
\providecommand \doibase [0]{http://dx.doi.org/}%
\providecommand \Doi [0]{\begingroup \@sanitize@url \@Doi }%
\providecommand \@Doi  [1]{\endgroup\@@startlink{\doibase#1}\@@Doi}%
\providecommand \@@Doi [1]{#1\@@endlink}%
\providecommand \selectlanguage [0]{\@gobble}%
\providecommand \bibinfo  [0]{\@secondoftwo}%
\providecommand \bibfield  [0]{\@secondoftwo}%
\providecommand \translation [1]{[#1]}%
\providecommand \BibitemOpen [0]{}%
\providecommand \bibitemStop [0]{}%
\providecommand \bibitemNoStop [0]{.\EOS\space}%
\providecommand \EOS [0]{\spacefactor3000\relax}%
\providecommand \BibitemShut  [1]{\csname bibitem#1\endcsname}%
%</preamble>
\bibitem [{\citenamefont {Novoselov}\ \emph {et~al.}(2004)\citenamefont
  {Novoselov}, \citenamefont {Geim}, \citenamefont {Morozov}, \citenamefont
  {Jiang}, \citenamefont {Zhang}, \citenamefont {Dubonos}, \citenamefont
  {Grigorieva},\ and\ \citenamefont {Firsov}}]{Novoselov22102004}%
  \BibitemOpen
  \bibfield  {author} {\bibinfo {author} {\bibfnamefont {K.~S.}\ \bibnamefont
  {Novoselov}}, \bibinfo {author} {\bibfnamefont {A.~K.}\ \bibnamefont {Geim}},
  \bibinfo {author} {\bibfnamefont {S.~V.}\ \bibnamefont {Morozov}}, \bibinfo
  {author} {\bibfnamefont {D.}~\bibnamefont {Jiang}}, \bibinfo {author}
  {\bibfnamefont {Y.}~\bibnamefont {Zhang}}, \bibinfo {author} {\bibfnamefont
  {S.~V.}\ \bibnamefont {Dubonos}}, \bibinfo {author} {\bibfnamefont {I.~V.}\
  \bibnamefont {Grigorieva}}, \ and\ \bibinfo {author} {\bibfnamefont {A.~A.}\
  \bibnamefont {Firsov}},\ }\Doi {10.1126/science.1102896} {\bibfield
  {journal} {\bibinfo  {journal} {Science},\ }\textbf {\bibinfo {volume}
  {306}},\ \bibinfo {pages} {666} (\bibinfo {year} {2004})},\ \Eprint
  {http://arxiv.org/abs/http://www.sciencemag.org/content/306/5696/666.full.pd%
f} {http://www.sciencemag.org/content/306/5696/666.full.pdf} \BibitemShut
  {NoStop}%
\bibitem [{\citenamefont {Castro~Neto}\ \emph {et~al.}(2009)\citenamefont
  {Castro~Neto}, \citenamefont {Guinea}, \citenamefont {Peres}, \citenamefont
  {Novoselov},\ and\ \citenamefont {Geim}}]{RevModPhys.81.109}%
  \BibitemOpen
  \bibfield  {author} {\bibinfo {author} {\bibfnamefont {A.~H.}\ \bibnamefont
  {Castro~Neto}}, \bibinfo {author} {\bibfnamefont {F.}~\bibnamefont {Guinea}},
  \bibinfo {author} {\bibfnamefont {N.~M.~R.}\ \bibnamefont {Peres}}, \bibinfo
  {author} {\bibfnamefont {K.~S.}\ \bibnamefont {Novoselov}}, \ and\ \bibinfo
  {author} {\bibfnamefont {A.~K.}\ \bibnamefont {Geim}},\ }\Doi
  {10.1103/RevModPhys.81.109} {\bibfield  {journal} {\bibinfo  {journal} {Rev.
  Mod. Phys.},\ }\textbf {\bibinfo {volume} {81}},\ \bibinfo {pages} {109}
  (\bibinfo {year} {2009})}\BibitemShut {NoStop}%
\bibitem [{\citenamefont {Gonz\'alez}\ \emph {et~al.}(2001)\citenamefont
  {Gonz\'alez}, \citenamefont {Guinea},\ and\ \citenamefont
  {Vozmediano}}]{PhysRevB.63.134421}%
  \BibitemOpen
  \bibfield  {author} {\bibinfo {author} {\bibfnamefont {J.}~\bibnamefont
  {Gonz\'alez}}, \bibinfo {author} {\bibfnamefont {F.}~\bibnamefont {Guinea}},
  \ and\ \bibinfo {author} {\bibfnamefont {M.~A.~H.}\ \bibnamefont
  {Vozmediano}},\ }\Doi {10.1103/PhysRevB.63.134421} {\bibfield  {journal}
  {\bibinfo  {journal} {Phys. Rev. B},\ }\textbf {\bibinfo {volume} {63}},\
  \bibinfo {pages} {134421} (\bibinfo {year} {2001})}\BibitemShut {NoStop}%
\bibitem [{\citenamefont {Kotov}\ \emph {et~al.}(2012)\citenamefont {Kotov},
  \citenamefont {Uchoa}, \citenamefont {Pereira}, \citenamefont {Guinea},\ and\
  \citenamefont {Castro~Neto}}]{RevModPhys.84.1067}%
  \BibitemOpen
  \bibfield  {author} {\bibinfo {author} {\bibfnamefont {V.~N.}\ \bibnamefont
  {Kotov}}, \bibinfo {author} {\bibfnamefont {B.}~\bibnamefont {Uchoa}},
  \bibinfo {author} {\bibfnamefont {V.~M.}\ \bibnamefont {Pereira}}, \bibinfo
  {author} {\bibfnamefont {F.}~\bibnamefont {Guinea}}, \ and\ \bibinfo {author}
  {\bibfnamefont {A.~H.}\ \bibnamefont {Castro~Neto}},\ }\Doi
  {10.1103/RevModPhys.84.1067} {\bibfield  {journal} {\bibinfo  {journal} {Rev.
  Mod. Phys.},\ }\textbf {\bibinfo {volume} {84}},\ \bibinfo {pages} {1067}
  (\bibinfo {year} {2012})}\BibitemShut {NoStop}%
\bibitem [{\citenamefont {Makogon}\ \emph {et~al.}(2011)\citenamefont
  {Makogon}, \citenamefont {van Gelderen}, \citenamefont {Rold\'an},\ and\
  \citenamefont {Smith}}]{PhysRevB.84.125404}%
  \BibitemOpen
  \bibfield  {author} {\bibinfo {author} {\bibfnamefont {D.}~\bibnamefont
  {Makogon}}, \bibinfo {author} {\bibfnamefont {R.}~\bibnamefont {van
  Gelderen}}, \bibinfo {author} {\bibfnamefont {R.}~\bibnamefont {Rold\'an}}, \
  and\ \bibinfo {author} {\bibfnamefont {C.~M.}\ \bibnamefont {Smith}},\ }\Doi
  {10.1103/PhysRevB.84.125404} {\bibfield  {journal} {\bibinfo  {journal}
  {Phys. Rev. B},\ }\textbf {\bibinfo {volume} {84}},\ \bibinfo {pages}
  {125404} (\bibinfo {year} {2011})}\BibitemShut {NoStop}%
\bibitem [{\citenamefont {Honerkamp}(2008)}]{PhysRevLett.100.146404}%
  \BibitemOpen
  \bibfield  {author} {\bibinfo {author} {\bibfnamefont {C.}~\bibnamefont
  {Honerkamp}},\ }\Doi {10.1103/PhysRevLett.100.146404} {\bibfield  {journal}
  {\bibinfo  {journal} {Phys. Rev. Lett.},\ }\textbf {\bibinfo {volume}
  {100}},\ \bibinfo {pages} {146404} (\bibinfo {year} {2008})}\BibitemShut
  {NoStop}%
\bibitem [{\citenamefont {Valenzuela}\ and\ \citenamefont
  {Vozmediano}(2008)}]{1367-2630-10-11-113009}%
  \BibitemOpen
  \bibfield  {author} {\bibinfo {author} {\bibfnamefont {B.}~\bibnamefont
  {Valenzuela}}\ and\ \bibinfo {author} {\bibfnamefont {M.~A.~H.}\ \bibnamefont
  {Vozmediano}},\ }\href {http://stacks.iop.org/1367-2630/10/i=11/a=113009}
  {\bibfield  {journal} {\bibinfo  {journal} {New Journal of Physics},\
  }\textbf {\bibinfo {volume} {10}},\ \bibinfo {pages} {113009} (\bibinfo
  {year} {2008})}\BibitemShut {NoStop}%
\bibitem [{\citenamefont {Black-Schaffer}\ and\ \citenamefont
  {Doniach}(2007)}]{PhysRevB.75.134512}%
  \BibitemOpen
  \bibfield  {author} {\bibinfo {author} {\bibfnamefont {A.~M.}\ \bibnamefont
  {Black-Schaffer}}\ and\ \bibinfo {author} {\bibfnamefont {S.}~\bibnamefont
  {Doniach}},\ }\Doi {10.1103/PhysRevB.75.134512} {\bibfield  {journal}
  {\bibinfo  {journal} {Phys. Rev. B},\ }\textbf {\bibinfo {volume} {75}},\
  \bibinfo {pages} {134512} (\bibinfo {year} {2007})}\BibitemShut {NoStop}%
\bibitem [{\citenamefont {Gonz\'alez}(2008)}]{PhysRevB.78.205431}%
  \BibitemOpen
  \bibfield  {author} {\bibinfo {author} {\bibfnamefont {J.}~\bibnamefont
  {Gonz\'alez}},\ }\Doi {10.1103/PhysRevB.78.205431} {\bibfield  {journal}
  {\bibinfo  {journal} {Phys. Rev. B},\ }\textbf {\bibinfo {volume} {78}},\
  \bibinfo {pages} {205431} (\bibinfo {year} {2008})}\BibitemShut {NoStop}%
\bibitem [{\citenamefont {Pathak}\ \emph {et~al.}(2010)\citenamefont {Pathak},
  \citenamefont {Shenoy},\ and\ \citenamefont {Baskaran}}]{PhysRevB.81.085431}%
  \BibitemOpen
  \bibfield  {author} {\bibinfo {author} {\bibfnamefont {S.}~\bibnamefont
  {Pathak}}, \bibinfo {author} {\bibfnamefont {V.~B.}\ \bibnamefont {Shenoy}},
  \ and\ \bibinfo {author} {\bibfnamefont {G.}~\bibnamefont {Baskaran}},\ }\Doi
  {10.1103/PhysRevB.81.085431} {\bibfield  {journal} {\bibinfo  {journal}
  {Phys. Rev. B},\ }\textbf {\bibinfo {volume} {81}},\ \bibinfo {pages}
  {085431} (\bibinfo {year} {2010})}\BibitemShut {NoStop}%
\bibitem [{\citenamefont {Loktev}\ and\ \citenamefont
  {Turkowski}(2009)}]{ScBosonExchange}%
  \BibitemOpen
  \bibfield  {author} {\bibinfo {author} {\bibfnamefont {V.~M.}\ \bibnamefont
  {Loktev}}\ and\ \bibinfo {author} {\bibfnamefont {V.}~\bibnamefont
  {Turkowski}},\ }\Doi {http://dx.doi.org/10.1063/1.3224719} {\bibfield
  {journal} {\bibinfo  {journal} {Low Temperature Physics},\ }\textbf {\bibinfo
  {volume} {35}},\ \bibinfo {pages} {632} (\bibinfo {year} {2009})}\BibitemShut
  {NoStop}%
\bibitem [{\citenamefont {McChesney}\ \emph {et~al.}(2010)\citenamefont
  {McChesney}, \citenamefont {Bostwick}, \citenamefont {Ohta}, \citenamefont
  {Seyller}, \citenamefont {Horn}, \citenamefont {Gonz\'alez},\ and\
  \citenamefont {Rotenberg}}]{PhysRevLett.104.136803}%
  \BibitemOpen
  \bibfield  {author} {\bibinfo {author} {\bibfnamefont {J.~L.}\ \bibnamefont
  {McChesney}}, \bibinfo {author} {\bibfnamefont {A.}~\bibnamefont {Bostwick}},
  \bibinfo {author} {\bibfnamefont {T.}~\bibnamefont {Ohta}}, \bibinfo {author}
  {\bibfnamefont {T.}~\bibnamefont {Seyller}}, \bibinfo {author} {\bibfnamefont
  {K.}~\bibnamefont {Horn}}, \bibinfo {author} {\bibfnamefont {J.}~\bibnamefont
  {Gonz\'alez}}, \ and\ \bibinfo {author} {\bibfnamefont {E.}~\bibnamefont
  {Rotenberg}},\ }\Doi {10.1103/PhysRevLett.104.136803} {\bibfield  {journal}
  {\bibinfo  {journal} {Phys. Rev. Lett.},\ }\textbf {\bibinfo {volume}
  {104}},\ \bibinfo {pages} {136803} (\bibinfo {year} {2010})}\BibitemShut
  {NoStop}%
\bibitem [{\citenamefont {Efetov}\ and\ \citenamefont
  {Kim}(2010)}]{PhysRevLett.105.256805}%
  \BibitemOpen
  \bibfield  {author} {\bibinfo {author} {\bibfnamefont {D.~K.}\ \bibnamefont
  {Efetov}}\ and\ \bibinfo {author} {\bibfnamefont {P.}~\bibnamefont {Kim}},\
  }\Doi {10.1103/PhysRevLett.105.256805} {\bibfield  {journal} {\bibinfo
  {journal} {Phys. Rev. Lett.},\ }\textbf {\bibinfo {volume} {105}},\ \bibinfo
  {pages} {256805} (\bibinfo {year} {2010})}\BibitemShut {NoStop}%
\bibitem [{\citenamefont {Nandkishore}\ \emph {et~al.}(2012)\citenamefont
  {Nandkishore}, \citenamefont {Levitov},\ and\ \citenamefont
  {Chubukov}}]{ChiralSuperconductivity}%
  \BibitemOpen
  \bibfield  {author} {\bibinfo {author} {\bibfnamefont {R.}~\bibnamefont
  {Nandkishore}}, \bibinfo {author} {\bibfnamefont {L.~S.}\ \bibnamefont
  {Levitov}}, \ and\ \bibinfo {author} {\bibfnamefont {A.~V.}\ \bibnamefont
  {Chubukov}},\ }\href@noop {} {\bibfield  {journal} {\bibinfo  {journal} {Nat
  Phys},\ }\textbf {\bibinfo {volume} {8}},\ \bibinfo {pages} {158} (\bibinfo
  {year} {2012})}\BibitemShut {NoStop}%
\bibitem [{\citenamefont {Wang}\ \emph {et~al.}(2012)\citenamefont {Wang},
  \citenamefont {Xiang}, \citenamefont {Wang}, \citenamefont {Wang},
  \citenamefont {Yang},\ and\ \citenamefont {Lee}}]{PhysRevB.85.035414}%
  \BibitemOpen
  \bibfield  {author} {\bibinfo {author} {\bibfnamefont {W.-S.}\ \bibnamefont
  {Wang}}, \bibinfo {author} {\bibfnamefont {Y.-Y.}\ \bibnamefont {Xiang}},
  \bibinfo {author} {\bibfnamefont {Q.-H.}\ \bibnamefont {Wang}}, \bibinfo
  {author} {\bibfnamefont {F.}~\bibnamefont {Wang}}, \bibinfo {author}
  {\bibfnamefont {F.}~\bibnamefont {Yang}}, \ and\ \bibinfo {author}
  {\bibfnamefont {D.-H.}\ \bibnamefont {Lee}},\ }\Doi
  {10.1103/PhysRevB.85.035414} {\bibfield  {journal} {\bibinfo  {journal}
  {Phys. Rev. B},\ }\textbf {\bibinfo {volume} {85}},\ \bibinfo {pages}
  {035414} (\bibinfo {year} {2012})}\BibitemShut {NoStop}%
\bibitem [{\citenamefont {Kiesel}\ \emph {et~al.}(2012)\citenamefont {Kiesel},
  \citenamefont {Platt}, \citenamefont {Hanke}, \citenamefont {Abanin},\ and\
  \citenamefont {Thomale}}]{PhysRevB.86.020507}%
  \BibitemOpen
  \bibfield  {author} {\bibinfo {author} {\bibfnamefont {M.~L.}\ \bibnamefont
  {Kiesel}}, \bibinfo {author} {\bibfnamefont {C.}~\bibnamefont {Platt}},
  \bibinfo {author} {\bibfnamefont {W.}~\bibnamefont {Hanke}}, \bibinfo
  {author} {\bibfnamefont {D.~A.}\ \bibnamefont {Abanin}}, \ and\ \bibinfo
  {author} {\bibfnamefont {R.}~\bibnamefont {Thomale}},\ }\Doi
  {10.1103/PhysRevB.86.020507} {\bibfield  {journal} {\bibinfo  {journal}
  {Phys. Rev. B},\ }\textbf {\bibinfo {volume} {86}},\ \bibinfo {pages}
  {020507} (\bibinfo {year} {2012})}\BibitemShut {NoStop}%
\bibitem [{\citenamefont {Ma}\ \emph {et~al.}(2011)\citenamefont {Ma},
  \citenamefont {Huang}, \citenamefont {Hu},\ and\ \citenamefont
  {Lin}}]{PhysRevB.84.121410}%
  \BibitemOpen
  \bibfield  {author} {\bibinfo {author} {\bibfnamefont {T.}~\bibnamefont
  {Ma}}, \bibinfo {author} {\bibfnamefont {Z.}~\bibnamefont {Huang}}, \bibinfo
  {author} {\bibfnamefont {F.}~\bibnamefont {Hu}}, \ and\ \bibinfo {author}
  {\bibfnamefont {H.-Q.}\ \bibnamefont {Lin}},\ }\Doi
  {10.1103/PhysRevB.84.121410} {\bibfield  {journal} {\bibinfo  {journal}
  {Phys. Rev. B},\ }\textbf {\bibinfo {volume} {84}},\ \bibinfo {pages}
  {121410} (\bibinfo {year} {2011})}\BibitemShut {NoStop}%
\bibitem [{\citenamefont {Wu}\ \emph {et~al.}(2013)\citenamefont {Wu},
  \citenamefont {Scherer}, \citenamefont {Honerkamp},\ and\ \citenamefont
  {Le~Hur}}]{PhysRevB.87.094521}%
  \BibitemOpen
  \bibfield  {author} {\bibinfo {author} {\bibfnamefont {W.}~\bibnamefont
  {Wu}}, \bibinfo {author} {\bibfnamefont {M.~M.}\ \bibnamefont {Scherer}},
  \bibinfo {author} {\bibfnamefont {C.}~\bibnamefont {Honerkamp}}, \ and\
  \bibinfo {author} {\bibfnamefont {K.}~\bibnamefont {Le~Hur}},\ }\Doi
  {10.1103/PhysRevB.87.094521} {\bibfield  {journal} {\bibinfo  {journal}
  {Phys. Rev. B},\ }\textbf {\bibinfo {volume} {87}},\ \bibinfo {pages}
  {094521} (\bibinfo {year} {2013})}\BibitemShut {NoStop}%
\bibitem [{\citenamefont {Nandkishore}\ and\ \citenamefont
  {Chubukov}(2012)}]{PhysRevB.86.115426}%
  \BibitemOpen
  \bibfield  {author} {\bibinfo {author} {\bibfnamefont {R.}~\bibnamefont
  {Nandkishore}}\ and\ \bibinfo {author} {\bibfnamefont {A.~V.}\ \bibnamefont
  {Chubukov}},\ }\Doi {10.1103/PhysRevB.86.115426} {\bibfield  {journal}
  {\bibinfo  {journal} {Phys. Rev. B},\ }\textbf {\bibinfo {volume} {86}},\
  \bibinfo {pages} {115426} (\bibinfo {year} {2012})}\BibitemShut {NoStop}%
\bibitem [{\citenamefont {Black-Schaffer}\ and\ \citenamefont
  {Doniach}(2010)}]{PhysRevB.81.014517}%
  \BibitemOpen
  \bibfield  {author} {\bibinfo {author} {\bibfnamefont {A.~M.}\ \bibnamefont
  {Black-Schaffer}}\ and\ \bibinfo {author} {\bibfnamefont {S.}~\bibnamefont
  {Doniach}},\ }\Doi {10.1103/PhysRevB.81.014517} {\bibfield  {journal}
  {\bibinfo  {journal} {Phys. Rev. B},\ }\textbf {\bibinfo {volume} {81}},\
  \bibinfo {pages} {014517} (\bibinfo {year} {2010})}\BibitemShut {NoStop}%
\bibitem [{\citenamefont {Black-Schaffer}(2013)}]{PhysRevB.88.104506}%
  \BibitemOpen
  \bibfield  {author} {\bibinfo {author} {\bibfnamefont {A.~M.}\ \bibnamefont
  {Black-Schaffer}},\ }\Doi {10.1103/PhysRevB.88.104506} {\bibfield  {journal}
  {\bibinfo  {journal} {Phys. Rev. B},\ }\textbf {\bibinfo {volume} {88}},\
  \bibinfo {pages} {104506} (\bibinfo {year} {2013})}\BibitemShut {NoStop}%
\bibitem [{\citenamefont {Black-Schaffer}(2012)}]{PhysRevLett.109.197001}%
  \BibitemOpen
  \bibfield  {author} {\bibinfo {author} {\bibfnamefont {A.~M.}\ \bibnamefont
  {Black-Schaffer}},\ }\Doi {10.1103/PhysRevLett.109.197001} {\bibfield
  {journal} {\bibinfo  {journal} {Phys. Rev. Lett.},\ }\textbf {\bibinfo
  {volume} {109}},\ \bibinfo {pages} {197001} (\bibinfo {year}
  {2012})}\BibitemShut {NoStop}%
\bibitem [{\citenamefont {Balatsky}\ \emph {et~al.}(2006)\citenamefont
  {Balatsky}, \citenamefont {Vekhter},\ and\ \citenamefont
  {Zhu}}]{RevModPhys.78.373}%
  \BibitemOpen
  \bibfield  {author} {\bibinfo {author} {\bibfnamefont {A.~V.}\ \bibnamefont
  {Balatsky}}, \bibinfo {author} {\bibfnamefont {I.}~\bibnamefont {Vekhter}}, \
  and\ \bibinfo {author} {\bibfnamefont {J.-X.}\ \bibnamefont {Zhu}},\ }\Doi
  {10.1103/RevModPhys.78.373} {\bibfield  {journal} {\bibinfo  {journal} {Rev.
  Mod. Phys.},\ }\textbf {\bibinfo {volume} {78}},\ \bibinfo {pages} {373}
  (\bibinfo {year} {2006})}\BibitemShut {NoStop}%
\bibitem [{\citenamefont {Anderson}(1959)}]{Anderson195926}%
  \BibitemOpen
  \bibfield  {author} {\bibinfo {author} {\bibfnamefont {P.}~\bibnamefont
  {Anderson}},\ }\Doi {http://dx.doi.org/10.1016/0022-3697(59)90036-8}
  {\bibfield  {journal} {\bibinfo  {journal} {Journal of Physics and Chemistry
  of Solids},\ }\textbf {\bibinfo {volume} {11}},\ \bibinfo {pages} {26 }
  (\bibinfo {year} {1959})},\ ISSN \bibinfo {issn} {0022-3697}\BibitemShut
  {NoStop}%
\bibitem [{\citenamefont {Baskaran}(2002)}]{PhysRevB.65.212505}%
  \BibitemOpen
  \bibfield  {author} {\bibinfo {author} {\bibfnamefont {G.}~\bibnamefont
  {Baskaran}},\ }\Doi {10.1103/PhysRevB.65.212505} {\bibfield  {journal}
  {\bibinfo  {journal} {Phys. Rev. B},\ }\textbf {\bibinfo {volume} {65}},\
  \bibinfo {pages} {212505} (\bibinfo {year} {2002})}\BibitemShut {NoStop}%
\bibitem [{\citenamefont {Baskaran}(2009)}]{TheFiveFoldWay}%
  \BibitemOpen
  \bibfield  {author} {\bibinfo {author} {\bibfnamefont {G.}~\bibnamefont
  {Baskaran}},\ }\Doi {10.1007/s12043-009-0094-8} {\bibfield  {journal}
  {\bibinfo  {journal} {Pramana},\ }\textbf {\bibinfo {volume} {73}},\ \bibinfo
  {pages} {61} (\bibinfo {year} {2009})},\ ISSN \bibinfo {issn}
  {0304-4289}\BibitemShut {NoStop}%
\end{thebibliography}
\end{document}